\documentclass[a4paper,12pt]{article}
\usepackage{multicol}

\title{A Theory
 of Measurement Uncertainty Based on 
Conditional Probability\footnote{Special contribution to the
session {\it ``The role and meaning of conditional probability
in Bayesian statistics''}, 
Section on Bayesian Statistical Science  
at the ``1996 Joint Statistical Meetings'',
Chicago, Illinois, August 4-8, 1996.\protect\newline
Email: dagostini@vaxrom.roma1.infn.it}
}

\author{  Giulio D'Agostini\\
Dip. di Fisica,
Universit\`a ``La Sapienza'' and INFN
(Italy)}

\date{}

\begin{document}
 \maketitle
\begin{abstract}
A theory of measurement uncertainty is presented, 
which, since it 
is based exclusively on the Bayesian approach and on the 
subjective concept 
of conditional probability, is applicable in the most general cases.

The recent International Organization for Standardization 
(ISO) 
recommendation on measurement uncertainty is reobtained as the 
limit case 
in which linearization is meaningful and one is interested only in 
the best 
estimates of the quantities and in their variances.
\end{abstract}
\mbox{}
\vspace{-13cm}
\begin{flushleft}
\tt 
Roma1 N.1079\newline
November 1996
\end{flushleft}
\vspace{13cm}
\subsection*{Introduction}
The value of a physical quantity obtained as a result of a 
measurement 
has a degree of uncertainty, due to unavoidable errors, of which 
one can 
recognize the source but never establish the exact magnitude. The 
uncertainty due to so called {\it statistical}
 errors is usually treated 
using the 
frequentistic concept of confidence intervals, although the 
procedure is 
rather unnatural and there are known cases (of great relevance in 
frontier 
research) in which this approach is not applicable. On the other 
hand, there 
is no way, within this frame, to handle uncertainties due to 
{\it systematic} errors 
in a consistent way.

Bayesian statistics, however, allows a theory of 
measurement 
uncertainty to be built which is applicable to all cases. The 
outcomes are in 
agreement with the recommendation of the Bureau International 
des Poids 
et Mesures (BIPM) and of the International Organization for the 
Standardization (ISO), which has also recognized the crucial role 
of 
subjective probability in assessing and expressing measurement 
uncertainty.

In the next section I will make some remarks about the 
implicit use in 
science of the intuitive concept of probability as degree of belief. 
Then I will 
briefly discuss the part of the BIPM recommendation which 
deals with 
subjective probability. The Bayesian theory of uncertainty which 
provides 
the mathematical foundation of the recommendation will be 
commented 
upon. Finally I will introduce an alternative theory, based 
exclusively on the 
Bayesian approach and on conditional probability. More details, 
including 
many practical examples, can be found in \cite{primer}.

\subsection*{Claimed frequentism versus practiced subjectivism}
Most physicists (I deal here mainly with Physics because 
of personal 
biases, but the remarks and the conclusions could easily be 
extended to 
other fields of research) have received a scientific education in 
which the 
concept of probability is related to the ratio of favorable over 
possible 
events, and to relative frequencies for the outcomes of repeated 
experiments.
Usually the first "definition'' ({\it combinatorial})
 is used in {\it theoretical} 
calculations and the second one ({\it frequentistic}) in {\it empirical} 
evaluations. The 
{\it subjective} definition of probability, as ``degree of belief'', is, 
instead, viewed 
with suspicion and usually misunderstood. The usual criticism is 
that 
``science must be objective'' and, hence that ``there should be no 
room for 
subjectivity''. Some even say: ``I do not \underline{believe} something. 
I  \underline{assess} it. This 
is not a matter for religion!''.

It is beyond the purposes of this paper to discuss the issue 
of the so 
called "objectivity'' of scientific results. I would just like to 
remind the 
reader that, as well expressed by the science historian 
Galison\cite{Galison},
\begin{quote}
{\sl \small
 ``Experiments begin and end in a matrix of beliefs. \ldots
 beliefs in instrument
types, in programs of experiment enquiry, in the trained, 
individual judgements
about every local behavior of pieces of apparatus \ldots''.}
\end{quote}

In my experience, and after interviewing many colleagues 
from several 
countries, physicists use (albeit unconsciously) the intuitive 
concept of 
probability as ``degree of belief", even for ``professional 
purposes''. 
Nevertheless, they have difficulty in accepting such a definition 
rationally, 
because - in my opinion - of their academic training. For 
example, apart 
from a small minority of orthodox frequentists, almost 
everybody accepts 
statements of the kind ``there is $90\,\%$
 probability that the value of 
the Top 
quark mass is between \ldots ". In general, in fact, even the 
frequentistic 
concept of {\it confidence interval} is usually interpreted in a 
\underline{subjective way}, 
and the correct statement (according to the frequentistic school) of 
``$90\,\%$ 
probability that the observed value lies in an interval around
$\mu$'' is 
usually 
{\it turned around} into a {\it ``$90\,\%$ 
\underline{probability} that
$\mu$ is around the 
observed value''}
($\mu$ indicates hereafter the true value).
 The reason is rather simple. 
A physicist 
- to continue with our example - seeks to obtain some knowledge 
about $\mu$
and, consciously or not, wants to understand which values of $\mu$
have high or 
low degrees of belief; or which intervals 
$\Delta \mu$ have large or small 
probability. 
A statement concerning the probability that a measured value falls 
within a 
certain interval around $\mu$ is sterile if it cannot be turned into an 
expression 
which states the quality of the knowledge of $\mu$ itself. 
Unfortunately, few 
scientists are aware that this can be done in a logically consistent 
way only 
by using the Bayes' theorem and some {\it \`a priori} degrees of belief. 
In practice, 
since one often deals with simple problems in which the 
likelihood is 
normal and the uniform distribution is a reasonable prior (in the 
sense that 
the same degree of belief is assigned to all the infinite values of $\mu$) 
the 
Bayes' formula is formally ``by-passed'' and the likelihood is 
taken as if it 
described the degrees of belief for $\mu$ after the outcome of the 
experiment is 
known (i.e. the final probability density function, if $\mu$ is a 
continuous 
quantity).
\subsection*{BIPM and ISO Recommendation on the measurement 
uncertainty}
An example which shows how this intuitive way of 
reasoning is so 
natural for the physicist can be found in the BIPM 
recommendation INC-1 
(1980) about the 
{\it ``expression of experimental uncertainty''}\cite{INC-1}.
 It  states that
\begin{quote}
{\sl\small
The uncertainty in the result of a measurement generally consists 
of 
several components which may be grouped into two categories 
according 
to the way in which their numerical value is estimated:
\begin{description}
\item[A:] those which are evaluated by statistical methods;
\item[B:] those which are evaluated by other means.
\end{description}
}
\end{quote}
Then it specifies that
\begin{quote}
{\sl\small 
The components in category B should be characterized by 
quantities 
$u_j^2$, which may be considered as approximations to the 
corresponding variances, the existence of which is assumed. The 
quantities $u_j^2$ may be treated like variances and the 
quantities $u_j$
like standard deviations.
}
\end{quote}
Clearly, this recommendation is meaningful only in a Bayesian 
framework. 
In fact, the recommendation has been criticized because it is not 
supported 
by conventional statistics (see e.g. \cite{Weise2} and references therein). 
Nevertheless, 
it has been approved and reaffirmed by the CIPM (Comit\'e
International des 
Poids et Mesures) and adopted by ISO in its {\it ``Guide to the 
expression of 
uncertainty in measurement''}\cite{ISO} and by NIST (National Institute 
of 
Standards and Technology) in an analogous guide\cite{nist}. In 
particular, the ISO 
{\it Guide} recognizes the crucial role of subjective probability in Type 
B 
uncertainties: 
\begin{quote}
{\sl \small
"\ldots  Type B standard uncertainty is obtained from an assumed 
probability
density function based on the degree of belief that an event will 
occur 
[often called subjective {\bf probability} \ldots].''
}
\end{quote}
\begin{quote}
{\sl \small

``Recommendation INC-1 (1980) upon which this Guide rests 
implicitly 
adopts such a viewpoint of probability ... as the appropriate way 
to 
calculate the combined standard uncertainty of a result of a 
measurement.''
}
\end{quote}
The BIPM recommendation and the ISO Guide deal only 
with 
definitions and with ``variance propagation'', performed, as usual, 
by 
linearization. A general theory has been proposed by Weise and 
W\"oger\cite{Weise2}. 
which they maintain should provide the mathematical foundation 
of the {\it Guide}. 
Their theory is based on Bayesian statistics and on the 
principle of 
maximum entropy. Although the authors show how powerful it 
is in many 
applications, the use of the maximum entropy principle is, in my 
opinion, a 
weak point which prevents the theory from being as general as 
claimed (see 
the remarks later on in this paper, on the choice of the priors) and 
which 
makes the formalism rather complicated.
	I show in the next section how it is possible to build an 
alternative 
theory, based exclusively on probability ``first principles'', which 
is very 
close to the physicist's intuition. In a certain sense the theory 
which will be 
proposed here can be seen as nothing more than a formalization 
of what 
most physicists unconsciously do.
\subsection*{A genuine Bayesian theory of measurement uncertainty}
In the Bayesian framework inference is performed by 
calculating the 
degrees of belief of the true values of the physical quantities, 
taking into 
account all the available information. Let us call
$\underline{x}=\{x_1, x_2, \ldots, x_{n_x}\} $
the 
{\it n-tuple} (``vector'') of observables, 
$\underline{\mu}=\{\mu_1, \mu_2, \ldots, \mu_{n_\mu}\}$
the n-tuple of the 
true values of the physical quantities of interest,  and 
$\underline{h}=\{h_1, h_2, \ldots, h_{n_h}\}$
the n-tuple of all the possible realizations of the {\it influence 
variables} $H_i$. The 
term ``influence variable'' is used here with an extended meaning, 
to indicate 
not only external factors which could influence the result 
(temperature, 
atmospheric pressure, etc.) but also any possible calibration 
constants and 
any source of systematic errors. In fact the distinction between 
$\underline{\mu}$ 
and $\underline{h}$ is 
artificial, since they are all 
{\it conditional hypotheses} for $\underline{x}$. We 
separate them 
simply because the aim of the research is to obtain knowledge 
about $\underline{\mu}$, 
while $\underline{h}$ are considered a nuisance.

The likelihood of the {\it sample} $\underline{x}$ 
being produced from $\underline{h}$ and 
$\underline{\mu}$ is
\begin{equation}
 f(\underline{x}|\underline{\mu}, \underline{h}, H_\circ)\,.
 \label{eq:cond}
 \end{equation}
$H_\circ$ is intended as a reminder that likelihoods and priors - and 
hence 
conclusions - depend on all explicit and implicit assumptions 
within the 
problem, and, in particular, on the parametric  functions used to 
model  
priors and likelihoods. (To simplify the formulae, $H_\circ$ will  no 
longer be 
written explicitly). Notice that (\ref{eq:cond})
 has to be meant as a function       
$f(\cdot|\underline{\mu},\underline{h})$
for all possible values of the sample $\underline{x}$, 
with no restrictions 
beyond those 
given by the coherence\cite{definetti}.

Using the Bayes' theorem we obtain, given an initial 
$f_\circ(\underline{\mu})$
which 
describes the different degrees of belief on all possible values of
$\underline{\mu}$
before 
the information on $\underline{x}$ 
is available, a final distribution $f(\underline{\mu})$
 for each 
possible set 
of values of the influence 
variables $\underline{h}$:
\begin{equation}
f(\underline{\mu}|\underline{x},\underline{h}) =
\frac{f(\underline{x}|\underline{\mu}, \underline{h})
      f_\circ(\underline{\mu})}
     {\int
      f(\underline{x}|\underline{\mu}, \underline{h})
       f_\circ(\underline{\mu})     
       d\underline{\mu}}\,.
\label{eq:ginf0}
\end{equation}

Notice that the integral over a probability density function (instead 
of a 
summation over discrete cases) is just used to simplify the 
notation. To 
obtain the final distribution of 
$\underline{\mu}$ one needs to re-weight 
(\ref{eq:ginf0}) with 
the degrees 
of belief on $\underline{h}$:
 \begin{equation}
f(\underline{\mu}|\underline{x}) =
\frac{\int f(\underline{x}|\underline{\mu}, \underline{h})
      f_\circ(\underline{\mu})f(\underline{h})d\underline{h}}
     {\int
      f(\underline{x}|\underline{\mu}, \underline{h})
       f_\circ(\underline{\mu})f(\underline{h})
       d\underline{\mu} d\underline{h}}\,.
\label{eq:ginf1}
\end{equation}
The same comment on the use of the integration, made after 
(\ref{eq:ginf0}), 
applies 
here. Although (\ref{eq:ginf1})
 is seldom used by physicists, the formula is 
conceptually equivalent to what experimentalists do when they 
vary all the 
parameters of the Monte Carlo simulation in order to estimate the 
``systematic error''\footnote{
Usually they are not interested in complete knowledge of 
$f(\underline{\mu})$
but only in best estimates and 
variances, and 
normality is assumed. Typical expressions one can find in 
publications, related to this procedure, are: 
``the 
following systematic checks have been performed'', and then 
``systematic errors have been added 
quadratically''.
}.

Notice that an alternative way of getting $f(\underline{\mu})$
 would be to 
first consider 
an initial joint probability density function 
$f_\circ(\underline{\mu}, \underline{h})$
and then to 
obtain $f(\underline{\mu})$ as 
the marginal of the final distribution 
$f(\underline{\mu}, \underline{h})$. Formula (\ref{eq:ginf1})
 is reobtained if $\underline{\mu}$ and $\underline{h}$ 
 are independent and if 
$f_\circ(\underline{\mu}, \underline{h})$
can be factorized into 
$f_\circ(\underline{\mu})$ and $f(\underline{h})$.
But this could be interpreted as an explicit requirement that 
$f(\underline{\mu}, \underline{h})$ exists, or 
even that the existence of $f(\underline{\mu}, \underline{h})$
 is needed for the assessment of 
 $f(\underline{x}|\underline{\mu},\underline{h})$.
As stated previously, 
$f(\underline{x}|\underline{\mu},\underline{h})$
simply describes the degree of belief 
on $\underline{x}$ for 
any conceivable configuration
$\{\underline{\mu},\underline{h}\}$, 
with no constraint other 
than 
coherence. This corresponds to what experimentalists do when 
they first 
give the result  with ``statistical uncertainty'' only and then look for 
all 
possible systematic effects and evaluate their related contributions 
to the 
``global uncertainty''.

\subsubsection*{Some comments about the choice of the priors}
I don't think that the problem of the prior choice is a 
fundamental issue. 
My view is that one should avoid pedantic discussions of the 
matter, 
because the idea of ``universally true priors'' reminds me terribly 
of the 
Byzanthine ``angels' sex'' debates.
If I had to give recommendations, they would be:
\begin{itemize}
\item
the {\it a priori} probability should be chosen in the same spirit 
as the rational 
person  who places a bet, seeking to minimize the risk of losing; 
\item
general principles may help, but, since it is difficult to 
apply elegant 
theoretical ideas to all practical situations, in many circumstances 
the 
{\it guess} of the ``expert'' can be relied on for guidance;
\item
in particular, I think - and in this respect I completely 
disagree with the 
authors of \cite{Weise2} - there is no reason why the maximum entropy 
principle 
should be used in an uncertainty theory, just because it is 
successful in 
statistical mechanics. In my opinion, while the use of this 
principle in 
the case of discrete random variables is as founded as Laplace's 
indifference principle, in the continuous case there exists the 
unavoidable problem of the choice of the right metric (``what is 
uniform 
in $x$ is not uniform in $x^2$''). It seems to me that the success of 
maximum 
entropy in statistical mechanics should be simply considered a 
lucky 
instance in which a physical scale (the Planck constant) provides 
the 
``right'' metrics in which the phase space cells are equiprobable.
\end{itemize}
In the following example I will use uniform and normal priors, 
which are 
reasonable for the problems considered.

\subsection*{An example: uncertainty due to unknown systematic error of the 
instrument 
scale offset}
In our scheme any influence quantity of which we do not 
know the 
exact value is a source of systematic error. It will change the final 
distribution of $\mu$ and hence its uncertainty. Let us take the case of 
the ``zero'' 
of an instrument, the value of which is never known exactly, due 
to limited 
accuracy and precision of the calibration. This lack of perfect 
knowledge can 
be modeled assuming that the zero {\it "true value''} $Z$ is normally 
distributed 
around 0 (i.e. the calibration was properly done!) with a standard 
deviation $\sigma_Z$.
 As far as $\mu$ is concerned, one may attribute the same degree 
of belief to 
all of its possible values. We can then take a uniform distribution 
defined 
over a large interval, chosen according to the characteristics of the 
measuring device and to our expectation on $\mu$. An alternative 
choice of 
vague priors could be a normal distribution with large variance 
and a 
reasonable average (the values have to be suggested by the best 
available 
knowledge of the measurand and of the experimental devices). 
For 
simplicity, a uniform distribution is chosen in this example.

As far as $f(x|\mu,z)$ is concerned, we may assume that, for 
all possible 
values of $\mu$ and $z$, the degree of belief for each value of the 
measured 
quantity $x$ can be described by a normal distribution with an 
expected value 
$\mu+z$ 
and variance $\sigma_\circ^2$:
\begin{equation}
f(x|\mu,z) = \frac{1}{\sqrt{2\pi}\sigma_\circ}
\exp{\left[-\frac{(x-\mu-z)^2}{2\sigma_\circ^2}\right]}\,.
\end{equation}
For each $z$ of the instrument offset we have a set of degrees of 
belief on $\mu$:
\begin{equation}
f(\mu|x,z) = \frac{1}{\sqrt{2\pi}\sigma_\circ}
\exp{\left[-\frac{(\mu-(x-z))^2}{2\sigma_\circ^2}\right]}\,.
\label{eq:fmuz}
\end{equation}
Weighting $f(\mu|z)$ with degrees of belief on $z$
 using (\ref{eq:ginf1}) we finally 
obtain
 \begin{equation}
f(\mu) \equiv f(\mu|x, \ldots,f_\circ(z)) =
\frac{1}{\sqrt{2\pi}\sqrt{\sigma_\circ^2+\sigma_Z^2}}
   \exp{\left[-\frac{(\mu-x)^2}{2(\sigma_\circ^2+\sigma_Z^2)}\right]}\,.
\end{equation}
The result is that $f(\mu)$  is still a gaussian, but with a 
variance larger than that due only to statistical effects. 
The 
global standard deviation is the quadratic combination of that due 
to the 
statistical fluctuation of the data sample and that due to the 
imperfect 
knowledge of the {\it systematic effect}:
\begin{equation}
\sigma_{tot}^2 = \sigma_\circ^2+\sigma_Z^2\,.
\end{equation}
This formula is well known and widely used, although nobody 
seems to 
care that it cannot be justified by conventional statistics.

It is interesting to notice that in this framework it makes 
no sense to 
speak of ``statistical'' and ``systematical'' uncertainties, as if they 
were of a 
different nature. They are all treated probabilistically. But this 
requires the 
concept of probability to be related to lack of knowledge, and not 
simply to 
the outcome of repeated experiments. This is in agreement with 
the 
classification in {\it Type A} and {\it Type B}
 of the components of the 
uncertainty, 
recommended by the BIPM.

If one has several sources of systematic errors, each 
related to an 
influence quantity, and such that their variations around their 
nominal values 
produce  linear variations to the measured value, then the ``usual'' 
combination of variances (and covariances) is obtained (see 
\cite{primer} 
for details).

	If several measurements are affected by the same 
unknown systematic 
error, their results are expected to be correlated. For example, 
considering 
only two measured values $x_1$ 
 and $x_2$ of the true values $\mu_1$ and 
$\mu_2$, 
the 
likelihood is
\begin{equation}
f(x_1,x_2|\mu_1,\mu_2,z) =
\frac{1}{2\pi\sigma_1\sigma_2}
\exp{\left[-\frac{1}{2}\left(
              \frac{(x_1-\mu_1-z)^2}{\sigma_1^2} +
              \frac{(x_2-\mu_2-z)^2}{\sigma_2^2}
                       \right)
     \right]}\,.
\nonumber
\end{equation}
The final distribution 
$f(\mu_1,\mu_2)$ is a bivariate normal distribution 
with 
expected values $x_1$ and $x_2$. The diagonal elements of the 
covariance matrix 
are $\sigma_i^2+\sigma_Z^2$, 
with $i=1,\ 2$. The covariance 
between $\mu_1$ and 
 $\mu_2$ is $\sigma_Z$ and their correlation factor is then
\begin{equation}
\rho(\mu_1, \mu_2)  = \frac{\sigma_Z^2}{\sqrt{\sigma_1^2+\sigma_Z^2}
                         \sqrt{\sigma_2^2+\sigma_Z^2}}\,.
\label{eq:rho1}
\end{equation}
The correlation coefficient is positively defined, as the definition 
of the 
systematic error considered here implies. Furthermore, as 
expected, several 
values influenced by the same unknown systematic error are 
correlated 
when the uncertainty due to the systematic error is comparable to 
- or larger 
than - the uncertainties due to sampling effects alone.
\subsection*{Conclusions}
Bayesian statistics is closer to the physicist's mentality and 
needs than 
one may na\"\i vely think. A Bayesian theory of measurement 
uncertainty has 
the simple and important role of formalizing what is often done, 
more or 
less intuitively, by experimentalists in simple cases, and to give 
guidance in 
more complex situations.
	
As far as the choice of the priors and the interpretation of 
conditional 
probability are concerned, it seems to me that, although it may 
look 
paradoxical at first sight, the "subjective'' approach ({\it \`a la} 
 de Finetti) has the 
best chance of achieving consensus among the scientific 
community (after 
some initial resistance due to cultural prejudices).

\end{document}